Structural Chemistry, Spin Order, and the Distinction between the Cuprate and Pnictide High-Temperature Superconductors

S. R. Ovshinsky
*Ovshinsky Innovation, 1050 East Square Lake Road, Bloomfield Hills, MI 48304*

In the cuprate and iron-pnictide systems, valence changes induce high-temperature superconductivity while the local structural chemistry and local spin order both independently generate the attractive interactions responsible for the high transition temperature. We argue that together they favor d-wave singlet superconductivity in the cuprates but s-wave singlet in the pnictides. This difference arises from the existence of a large on-site repulsion between carriers in the cuprates largely absent in the pnictides. Fluorine is responsible for raising $T_c$ significantly in some pnictides and in the cuprates to 155K-168K, the highest achieved at ambient pressure. We propose an experimental procedure for finding and fabricating the fluorinated cuprate phase having that exceptional property.

PACS numbers: 74.70.-b, 74.72.-h, 74.90.+n

In this paper we show that elementary considerations of crystal chemistry and local spin ordering, similar except for detail in the cuprate and iron-pnictide high temperature superconductors, sharply delimit the types of superconducting order particular to each class and explain the special role of fluorine substitution in dramatically raising the transition temperature in both classes of materials. These considerations strongly favor s-wave superconductivity in the pnictides in contrast to the d-wave superconductivity they favor in the cuprates.

No consensus has emerged as to the mechanism underlying high temperature superconductivity in copper-based materials. Recently, with the discovery of high temperature superconductivity in the ROFeAs family of materials (with R a rare earth) upon fluorine substitution for oxygen [1-4], a similar surge of diverse proposals for the mechanism has emerged, also with no consensus. It has been stated that with no clear present understanding of the Cu-based superconductors, how can one expect so quickly to understand the Fe-based materials [5]. It is the purpose of this note to point to parallels between them by describing two pairing mechanisms common to both which operate in the two families with important differences of detail. Our focus is on the structural chemistry and spin correlations giving rise to them in both families of materials, achieving thereby a very simple picture for understanding the superconducting order parameters of both classes of materials. The focus on structural chemistry allows us to pinpoint the specific difference between the two classes of materials responsible for the difference in their superconducting order parameters. While other pnictide systems have been discovered with transition temperatures as high as 37K, cogently summarized in [6], for brevity and clarity we confine our explicit discussion to the ROFeAs family. Considerations similar to those developed here for the ROFeAs family apply to the other pnictides. The picture developed here for the ROFeAs and the cuprates also explains the role of fluorine in both cases.

Without question, this picture is highly oversimplified. We argue only qualitatively and consider only the electronic structures of the individual Cu and Fe ions in their local environments, as in ligand field theory. Initially we ignore the broadening of the resulting energy levels into energy bands and the dynamic interplay between the band structure and the quasiparticle interactions which is central to any complete theory of superconductivity. Our justification is that the dominant features of the electronic structures are given correctly by such structural chemical considerations, as is well established for the cuprates. Moreover, going more deeply hasn't led to a consensus theory of the superconductivity of these materials. In the present paper, we argue that our consideration of the structural chemistry first does yield insights valuable for the construction of more complete theories.

The essential importance of the structural chemistry for understanding the superconductivity of the copper-based materials was first pointed out in 1987 [7]. The Cu ions in the $CuO_2$ planes of the parent materials are in the $Cu^{II}$ state with a $d^9$ configuration. Each has square – pyramidal coordination with 5 neighboring $O^{2-}$ ions. As a consequence the $d(x^2-y^2)$-state is half occupied, and the $Cu^{II}$ ion moves out of the plane towards the apical oxygen ion to reduce the repulsion of the $d(x^2-y^2)$ state by the in-plane oxygens. The parent materials, e.g. $RBaCu_3O_7$, are antiferromagnetic, with neighboring Cu ions in the $CuO_2$ plane having opposite spin orientations. We have pointed out [7], that removal of O atoms from the CuO lines results in ordered rows of O vacancies which act as acceptors, introducing holes into the Cu $d(x^2-y^2)$ levels in the $CuO_2$ planes. The introduction of a hole converts a $Cu^{II}$ state locally into a $Cu^{III}$ state. More specifically, the local $d(x^2-y^2)$ state is emptied, eliminating the repulsion between the previously occupied orbital and the neighboring oxygen ions, which results in a movement of the $Cu^{III}$ ion back towards the plane of the oxygens, *back towards* the ideal square pyramidal configuration. As the dihedral angle of the original $OCu^{II}O$ bonds is only 165°, this can be a very substantial structural change.

Hybridization between the Cu $d(x^2-y^2)$ orbital and the O $p_{x,y}$ orbitals allows the hole to move [8,9]. Each hole is thus accompanied as it moves by the above substantial structural distortion associated with the propagating valence change. Overlap of the structural distortions associated with two neighboring holes reduces the energy of distortion, resulting in an attractive pair interaction strongest for nearest neighbor holes and with a long elastic tail. We have proposed this valence-change-based interaction (structural-chemistry-based interaction) as a part of the interaction responsible for superconductivity [8,9].

The Cu $d(x^2-y^2)$ and O $p_{x,y}$ hybridization introduces a superexchange interaction responsible for the antiferromagnetic order in the parent compounds. The introduction of an increasing number of holes destroys the antiferromagnetic long-range order of the parent compounds, but, as we have emphasized [9], short-range antiferromagnetic order persists into the superconducting concentration range. Thus the holes have a spin even though the $Cu^{III}$ configuration does not, and the local short-range order of the background medium implies that there is a strong attraction of holes of antiparallel spin. However, two holes cannot be on the same site unless one hole is in a deeper level, which requires a substantial energy increase. There is, in effect, a strong on-site repulsion between holes. Acting together, the on-site repulsion, the near neighbor attraction, and the underlying local spin order strongly favor d-wave, spin-singlet superconductivity in which the probability of finding two holes on the same site vanishes.

In sum, then, we have proposed that the attractive interaction driving superconductivity in the cuprates has two components, one from the structural distortions associated with valence change and one associated with the underlying antiferromagnetic short-range order. We have proposed that these interactions are strong enough so that bound pairs of holes of opposite spin persist as bosons above the transition temperature [10]. The result is a non-Fermi-liquid state with mixed conduction by bosons at the Fermi level together with conventional quasiparticles.

We argue here that these same two sources of the attractive interaction driving superconductivity exist in the iron pnictides as well where, however, they reinforce. The Fe ions in the FeAs layers are in the $Fe^{II}$ state with a $d^6$ configuration in ROFeAs. Each has distorted tetrahedral coordination with the four neighboring $As^{3-}$ ions with two distinct Fe-As bond angles [1-4]. In PrOFeAs, the tetrahedron is squashed substantially along the c -axis so that its effective c/a ratio is 0.954 [11]. As a consequence of the near tetragonal structure, the $d(x^2–y^{2)}$ and $d(z^2)$ levels lie lowest and are occupied by four spin-paired electrons. The $d_{xy}$ level lies below the degenerate $d_{xz}$ and $d_{yz}$ levels by an amount $\Delta$. It is known [10] that the Fe spin is 1. Thus the $d_{xy}$ and one of the $d_{xz}$ and $d_{yz}$ orbitals must each be singly occupied with the two spins parallel. This implies that $\Delta < U – U' + J$, where U is the direct Coulomb interaction between two electrons of opposite spin in the same d orbital, U′ is the direct Coulomb interaction between two electrons in different d orbitals, and J is the exchange interaction between two electrons of parallel spins in different d orbitals. The superexchange between nearest neighbor spin-1 $Fe^{II}$ ions is antiferromagnetic , leading to a checkerboard spin ordering in each FeAs layer [10].

The parent compounds are low-temperature superconductors with a normal-state resistivity typical of semimetals [2]. Superconductivity is achieved by substituting fluorine for oxygen in the rare-earth oxide layers [1,2]. One effect of the fluorine substitution is the donation of electrons, one per F atom, to the FeAs layer. The result is a reduction, a local valence change from $Fe^{II}$ to $Fe^{I}$ with further occupancy of either the $d_{xy}$ orbitals or one of the $d_{xz}$ or $d_{yz}$ orbitals. If the former occurs, the spin is necessarily opposite to the spin direction of the $Fe^{II}$ state, changing the spin locally from spin 1 to spin ½. If the latter occurs, the spin is necessarily parallel to the spin direction of the $Fe^{II}$ state, by Hund's rule. We can infer that the former is the case from the fact that fluorine substitution ultimately destroys the long-range antiferromagnetic order. If the latter were the case, the superexchange interactions would have been strengthened and not reduced as in the former case, increasing the Neel temperature instead of destroying the long-range order. Now the requirement for $Fe^{I}$ to favor the spin ½ configuration over the spin 1 configuration is that $\Delta > U – U' + J$, in apparent contradiction to our conclusion about the $Fe^{II}$ state. However, this $Fe^{II}$ to $Fe^{I}$ valence change decreases the strength of the attractive Coulomb interaction between the Fe ion and its nearest neighbor $As^{3-}$ ions, which consequently move outward in the ab-plane, increasing the local effective c/a ratio and increasing $\Delta$ in turn from its original value $\Delta$ to $\Delta' > U - U' + J$. As these latter three quantities are all in the eV range, we can infer that the resulting distortion from adding an electron is substantial. This substantial *increase* in the distortion from ideal tetrahedral geometry is opposite to what happens in the cuprates, as is the reduction opposite to the oxidation from $Cu^{II}$ to $Cu^{III}$. Nevertheless, this substantial distortion similarly provides the basis for an attractive interaction between two neighboring electrons of significant strength for close pairs and with an elastic tail.

Similarly, the underlying antiferromagnetic order of the parent compounds introduces an additional attractive spin-based pairing interaction. However, having two extra electrons on the same Fe site no longer requires excitation to a higher state, as there is room for one more electron in the $d_{xz}$, $d_{yz}$ multiplet. There is a significantly lesser energetic cost and so a weaker on–site repulsion between carriers. Considering the structural-chemistry-based interaction alone, the superconducting order parameter would be an s-wave spin singlet to take full advantage of that interaction. The spin-based interaction is also optimized for an s-wave, spin-singlet order parameter, which is consistent with the underlying antiferromagnetic short-range order though less so with the preferred spin-1 configuration of the neutral Fe at the center of the pair wave function, implying a small additional on-site repulsion. We have supposed that the radius of the short-range spin order is comparable to or larger than the superconducting pair radius. Chen, et al. have found that $SmFeAsO_{0.85}F_{0.15}$ has a BCS-like s-wave spin-singlet order parameter [12], readily understood on the basis of these simple arguments.

Thus, in both classes of materials, high-temperature superconductivity arises from strong attractive interactions between carriers generated by the local structural-chemical changes associated with valence changes, $Cu^{II}$ to $Cu^{III}$ in the cuprates and $Fe^{II}$ to $Fe^{I}$ in the pnictides, and from the accompanying changes in the local spin order. Strong on-site repulsions militate against s-wave superconductivity in the cuprates, forcing the superconducting order parameter to become more complex, but not in the pnictides. This pinpoints the key distinction between the two classes of materials, the fact that the on-site repulsion of two electrons in the pnictides is substantially less than that between two holes in the cuprates, our most significant result.

Another strong parallel lies in the role of fluorine. In each class, fluorine addition has a profound effect on its superconductivity. In the RO-pnictides, substitution of oxygen by fluorine creates a carrier increase and is thus far the only source of superconductivity. In the cuprates substitution of oxygen by fluorine has another important function. In addition to providing carriers to the $CuO_2$ planes, we have found that it raises the transition temperature up to 155-168K as demonstrated by two different fluorine insertion methods [13-16], temperatures higher than those achieved under pressure [15]. In addition to providing carriers, fluorine contracts the entire lattice structure of the cuprates essentially isotropically, thereby increasing the transition temperature. In the RO-pnictides, that contraction and accompanying transition-temperature increase is achieved for given F concentration by incorporation of smaller rare earth ions [1-4,17].

Confirmation of our findings of 155-168K values for $T_c$ in fluorinated cuprates [12-15] was reported by three different groups, two in conference proceedings [18,19] and one in an archival journal [20]. Nevertheless, despite the potential importance of achieving such high transition temperatures at ambient pressures, there has been no follow up in the literature. This failure should not be understood as implying that the reports of refs.[13-16,18-20] should be disregarded. We believe that the results reported could have been intrinsically difficult to reproduce for a solid physical reason that can be inferred directly from the findings in refs.[13-16]. While a very clear signal of superconductivity was seen in the resistance, the Meissner effect was very weak. The samples were multiphase, with the 155-168K $T_c$'s observed only in the presence of a $YBa_2Cu_3O_{7-x}$ (YBCO) phase with very small F content [19]. Taken together, these facts imply that the 155-168K superconducting component was a fluorine-rich matrix phase associated with the boundaries of the YBCO grains. Judging from the magnitude of the Meissner effect, its volume fraction was very low, though large enough to form a continuous path through the sample, i. e. to exceed the percolation threshold for that morphology. Because a $BaF_2$ phase competed more successfully for the fluorine than the YBCO phase, it is quite possible that failure to observe the 155-168K transition merely implies that the percolation threshold had not been exceeded. However, there are currently available experimental techniques not available two decades ago which could overcome these difficulties. We propose the following experimental procedure for a systematic search for the fluorinated high $T_c$ phase: 1.) Create a fluorinated sample by rigorously following the procedures of refs. [13-16]. 2.) Irrespective of whether the 155-168K transition is observed (the sample may be below the percolation threshold), use a scanning tunneling microscope to search for the nanoscale regions of highest superconducting energy gap. 3.) If the maximum gap corresponds to the high temperatures observed in refs. [13-16], use structural and compositional nanoprobes to identify the phase responsible. 4.) Identify a substrate suitable for fabrication of a film of this candidate phase by layer-by-layer deposition techniques such as pulsed-laser deposition. 5.) Fabricate and test the resulting homogeneous samples for superconductivity.

I acknowledge very helpful discussions with M. H. Cohen and H. Fritzsche. This paper has been posted on the Cornell University preprint archive (arXiv) as number 0807.1673.

[1] Y. Kamihara, H. Hiramatsu, M. Hirano, R. Kawamura, H. Yanagi, T, Kamiya and H. Hosono, J. Am. Chem. Soc. 128 (2006) 10012.
[2] Y. Kamihara, T. Watanabe, M. Hirano and H. Hosono, J. Am. Chem Soc. 130 (2008) 3296.
[3] H. Takahashi, K. Arii, Y. Kamihara, M. Hirano and H. Hosono, Nature 453 (2008) 376.
[4] Z-A. Ren, W. Lu, J. Yang, W. Yi, X-L. Shen, Z-C. Li, G-C. Che, X-L. Dong, L-L. Sun, F. Zhou and Z-X. Zhao, Chin. Phys. Lett. 2008 25 (6):2215-2216.
[5] A. Cho, Science 320 (2008) 870.
[6] J. H. Tapp, Z. Tang, B. Lv, K. Sazmal, B. Lorenz, P. C. W. Chu, and A. M. Guloy, Phys. Rev. B78 (2008) 060505(R).
[7] S. R. Ovshinsky, S. J. Hudgens, R. L. Lintveldt, and D. B. Rorabacher, Modern Phys. Lett. B1 (1987) 275.
[8] S. R. Ovshinsky, Chem. Phys. Lett. 195 (1992) 455.
[9] S. R. Ovshinsky, Appl. Superconductivity 1 (1993) 263
[10] C. de la Cruz, Q.Huang, J. W. Lynn, J. Li, W. Ratcliff II, J. L. Zaretsky, H. A. Mook, G. F. Chen, J. L. Luo, N. L. Wang, and P. Dai, Nature 453 (2008) 899
[11] P. Quebe, L. J. Terbüchte, W. Jeitschko, J. Alloys and Comp, (2000) 302, 70

[12] T. Y. Chen, Z. Tesanovic, R. H. Liu, X. H. Chen, and C. L. Chien, Nature 453 (2008) 1224
[13] S. R. Ovshinsky, R. T. Young, D. D. Allred, G. DeMaggio and G.A. Van der Leeden, Phys. Rev. Lett. 58 (1987) 2579.
[14] S. R. Ovshinsky, R. T. Young, B. S. Chao, G. Fournier and D. A. Pawlik, Reviews of Solid State Sci. 1 (1987) 207.
[15] R. T. Young, S. R. Ovshinsky, B. S. Chao, G. Fournier and D.A. Pawlik, MRS Symp. Proc. 99 (1988) 349.
[16] L. Guo, Y. Y. Xue, F. Chen, Q. Xiong, P. L. Meng, D. Ramirez, C. W. Chu, J. H. Eggert, and H. K. Mao, Phys. Rev. B 50 (1994) 4260
[17] X. H. Chen, T. Wu, R. H. Liu, H, Chen, and D. F. Fang, Nature 453 (2008) 761
[18] Y. R. Meng, Y. R. Ren, M. Z. Lin, Q. Y. Tu, Z. J. Lin, L. H. Sang, W. Q. Bing, Proc. Int. Symp. on High $T_C$ Superconductors (1987), Beijing, China
[19] J. H. Kung, Proc. Symp on Low Temp. Phys (1987), Hsin-Chu, Taiwan
[20] R. N. Bargava, S. P. Herko, and W. N. Osborne, Phys. Rev. Lett. (1987), 1468